\documentclass[sigplan,screen,nonacm]{acmart}

\usepackage{amsmath}
\usepackage{tlatex}
\usepackage{microtype}
\usepackage{calc}
\usepackage{balance}

\newcommand{\mytodo}[1]{\noindent{\textcolor{red}{TODO: #1}}}
\newcommand{\tlatexnum}[2]{$\langle #1 \rangle #2$}

\title{Formal Verification of a Distributed Dynamic Reconfiguration Protocol}

\setcopyright{acmlicensed}
\acmPrice{15.00}
\acmDOI{10.1145/3497775.3503688}
\acmYear{2022}
\copyrightyear{2022}
\acmSubmissionID{poplws22cppmain-p30-p}
\acmISBN{978-1-4503-9182-5/22/01}
\acmConference[CPP '22]{Proceedings of the 11th ACM SIGPLAN International Conference on Certified Programs and Proofs}{January 17--18, 2022}{Philadelphia, PA, USA}
\acmBooktitle{Proceedings of the 11th ACM SIGPLAN International Conference on Certified Programs and Proofs (CPP '22), January 17--18, 2022, Philadelphia, PA, USA}

\titlenote{This work has been partially supported by NSF award CNS-1801546.}

\begin{document}

\author{William Schultz}
\authornote{Both authors contributed equally to this research.}
\affiliation{
    \institution{Northeastern University}
    \city{Boston}
    \state{MA}
    \country{USA}
}
\email{schultz.w@northeastern.edu}

\author{Ian Dardik}
\affiliation{
    \institution{Northeastern University}
    \city{Boston}
    \state{MA}
    \country{USA}
}
\email{dardik.i@northeastern.edu}
\authornotemark[2]

\author{Stavros Tripakis}
\affiliation{
    \institution{Northeastern University}
    \city{Boston}
    \state{MA}
    \country{USA}
}
\email{stavros@northeastern.edu}

\begin{abstract}
    We present a formal, machine checked TLA+ safety proof of \emph{MongoRaftReconfig}, a distributed dynamic reconfiguration protocol. \emph{MongoRaftReconfig} was designed for and implemented in MongoDB, a distributed database whose replication protocol is derived from the Raft consensus algorithm. We present an inductive invariant for \emph{MongoRaftReconfig} that is formalized in TLA+ and formally proved using the TLA+ proof system (TLAPS). We also present a formal TLAPS proof of two key safety properties of \textit{MongoRaftReconfig}, \textit{LeaderCompleteness} and \textit{StateMachineSafety}. To our knowledge, these are the first machine checked inductive invariant and safety proof of a dynamic reconfiguration protocol for a Raft based replication system.
\end{abstract}



\maketitle

\newcommand{\statloctlapsproofs}{$3189$}
\newcommand{\statlocindinv}{140}
\newcommand{\statlocprotspec}{359}
\newcommand{\statnumprooflemmas}{78}
\newcommand{\statnumproofthms}{3}

\section{Introduction}

Distributed replication systems based on the replicated state machine model \cite{Schneider1990} have become ubiquitous as the foundation of modern, fault-tolerant data storage systems. In order for these systems to ensure availability in the presence of faults, they must be able to dynamically replace failed nodes with healthy ones, a process known as \emph{dynamic reconfiguration}. 

The protocols for building distributed replication systems have been well studied and implemented in a variety of systems \cite{chandra2007paxos,corbett2013spanner,Huang2020,taft2020cockroachdb}. Paxos \cite{Lamport1998} and, more recently, Raft \cite{ongaro2014search}, have served as the logical basis for building provably correct distributed replication systems. Dynamic reconfiguration, however, is an additionally challenging and subtle problem \cite{aguilera2010} for the protocols underlying these systems. 

Furthermore, few of these reconfiguration protocols have been formally verified \cite{shraer2012dynamic, moraru2013there, ports2015designing}.
The Raft consensus protocol, originally published in 2014, provided a dynamic reconfiguration algorithm in its initial publication, but did not include a precise discussion of its correctness or include a formal specification or proof. A critical safety bug \cite{ongaro2015-membership-bug} in one of its reconfiguration protocols was found after initial publication, demonstrating that the design and verification of reconfiguration protocols for these systems is a challenging task. This also demonstrates that formal verification is valuable for ensuring correctness of these protocols.

MongoDB \cite{mongodb-repo} is a general purpose, document oriented database which implements a distributed replication system \cite{schultz2019tunable} for providing high availability and fault tolerance. MongoDB's replication system uses a log-based consensus protocol that derives from Raft \cite{zhou2021fault}. MongoDB recently introduced a novel dynamic reconfiguration protocol, \textit{MongoRaftReconfig}, for its replication system. The \textit{MongoRaftReconfig} protocol is described in detail in \cite{schultz2021design}, which includes a TLA+ formal specification of the protocol and a manual safety proof. 

In this paper, we present the first formal verification of the safety properties of \textit{MongoRaftReconfig}. We present a formally stated inductive invariant for the protocol, which we prove and then utilize to establish two high level safety properties of the protocol. In particular, we prove (1) \textit{LeaderCompleteness}, which, intuitively, states that if a log entry is committed it is durable, and (2) \textit{StateMachineSafety}, which says that log entries committed at a particular index must be consistent across all nodes in the system. We carry out our verification efforts using TLAPS, the TLA+ proof system \cite{cousineau2012tla}. To our knowledge, this is the first machine checked inductive invariant and safety proof of a reconfiguration protocol for a Raft based replication system. 



To summarize, we make the following contributions:
\begin{itemize}
    \item A formally stated TLA+ inductive invariant for the \textit{MongoRaftReconfig} protocol. To our knowledge, this is both the first inductive invariant for a Raft-based reconfiguration protocol and the first that has been formalized.
    
    \item A formally verified TLAPS proof of our inductive invariant.

    \item A formally verified TLAPS proof that \textit{MongoRaftReconfig} satisfies the two above safety properties, \textit{LeaderCompleteness} and \textit{StateMachineSafety}. To our knowledge, this is the first machine checked safety proof of a Raft-based reconfiguration protocol.
\end{itemize}
All of our TLA+ specifications, TLAPS proof code, and instructions for checking our proofs are included in the supplementary material \cite{zenodo-tlaps-safety-proof} for this paper. Where appropriate throughout this paper, we cite the relevant files located in this material.




The rest of this paper is organized as follows. In Section \ref{sec:background} we provide some general background about MongoDB replication and TLA+. Section \ref{sec:verification-problem-statement} provides a formal statement of the verification results that we establish in this paper. Section \ref{sec:inductive-invariant} presents our inductive invariant for \textit{MongoRaftReconfig}, and Section \ref{sec:tlaps-proof} presents our formal safety proof of \textit{MongoRaftReconfig} in TLAPS, which makes use of our inductive invariant. Section \ref{sec:related-work} discusses related work, and Section \ref{sec:conclusions} discusses conclusions and future work. 

\section{Background}
\label{sec:background}

\subsection{The MongoDB Static Replication Protocol}

MongoDB is a general purpose, document oriented database that stores data in JSON-like objects. A MongoDB database consists of a set of collections, where a collection is a set of unique documents. To provide high availability, MongoDB provides the ability to run a database as a \textit{replica set}, which is a set of MongoDB servers that act as a consensus group, where each server maintains a logical copy of the database state. 

MongoDB replica sets utilize a replication protocol that is derived from Raft \cite{OngaroDissertation2014}, with some extensions. We refer to MongoDB's abstract replication protocol, without reconfiguration, as \textit{MongoStaticRaft}, to distinguish it from the \textit{MongoRaftReconfig} protocol verified in this paper. \textit{MongoStaticRaft} can be viewed as a modified version of standard Raft that satisfies the same underlying correctness properties, and it is described in more detail in \cite{zhou2021fault,schultz2019tunable}. We provide a high level overview here, since \textit{MongoRaftReconfig} is built on top of \textit{MongoStaticRaft}. 

A MongoDB replica set running \textit{MongoStaticRaft} consists of a set of server processes, $Server = \{s_1,s_2,\dots,s_n\}$. There exists a single \textit{primary} server and a set of \textit{secondary} servers. As in standard Raft, there is a single primary elected per term. The primary server accepts client writes and inserts them into an ordered operation log known as the \textit{oplog}. The oplog is a logical log where each entry contains information about how to apply a single database operation. Each entry is assigned a monotonically increasing timestamp, and these timestamps are unique and totally ordered within a server log. These log entries are then replicated to secondaries which apply them in order leading to a consistent database state on all servers. When the primary learns that enough servers have replicated a log entry in its term, the primary will mark it as \textit{committed}, guaranteeing that the entry is permanently durable in the replica set.

\subsection{The MongoDB Dynamic Reconfiguration Protocol: MongoRaftReconfig}

\textit{MongoRaftReconfig}, the protocol verified in this paper, is an extension of \textit{MongoStaticRaft} that allows for dynamic reconfiguration. \textit{MongoRaftReconfig} utilizes a logless approach to managing configuration state and decouples the processing of configuration changes from the main database operation log. The full details of \textit{MongoRaftReconfig} are presented in \cite{schultz2021design}, but we provide a high level overview here. In \textit{MongoRaftReconfig}, each server of a replica set maintains a single, durable \textit{configuration}, where a configuration is formally defined as a tuple $(m,v,t)$, where $m \in 2^{Server}$ is a subset of all servers, $v \in \mathbb{N}$ is a numeric configuration \textit{version}, and $t \in \mathbb{N}$ is the numeric \textit{term} of the configuration. Configurations are totally ordered by their $(version, term)$ pair, where $term$ is compared first, followed by version. Servers can install any configuration newer than their own. Reconfiguration operations, which can only be processed by primary servers, update a server's local configuration to a new configuration specified by the client.

Prior to the work presented in this paper, \cite{schultz2021design} presented a formal TLA+ specification of \textit{MongoRaftReconfig}, results from model checking its safety on finite protocol instances, and a manual, prose safety proof. There existed, however, no formal inductive invariant or machine checked proof for the safety properties of \textit{MongoRaftReconfig}. The formal inductive invariant we present in this paper bears some structural similarity to the manual proof given in \cite{schultz2021design}, but the TLAPS proofs presented in this paper were developed independently, and are not based on the prior, manual proof. 


\subsection{TLA+ and TLAPS}
TLA+ \cite{lamport2002specifying} is a formal specification language for describing distributed and concurrent systems that is based on first order and temporal logic  \cite{Pnueli1977}. Since \emph{MongoRaftReconfig} is formally specified using the TLA+ language and it is the language used for our proofs, we provide a brief overview of TLA+ and its associated proof system, TLAPS \cite{cousineau2012tla}.

\subsubsection{Specifications in TLA+}

Specifying a system in TLA+ consists of defining a set of state variables, $vars$, along with a temporal logic formula which describes the set of permitted system behaviors over these variables. The canonical way of defining a specification is as the conjunction of an initial state predicate, $Init$, and a next state relation, $Next$, which determine, respectively, the set of allowed initial states and how the protocol may transition between states. The overall system is then defined by the temporal formula $Init \wedge \square [Next]_{vars}$, where $\square$ denotes the ``always" operator of temporal logic, meaning that a formula holds true at every step of a behavior, and $vars$ denotes a sequence of all state variables of a specification. $[Next]_{vars}$ is equivalent to the expression $Next \vee (vars' = vars)$, which means that specifications of this form allow for \textit{stuttering} steps i.e. transitions that do not change the state.  A primed TLA+ expression containing state variables, expressed by attaching a $'$ symbol, denotes the value of that expression in the next state of a system behavior. The next state relation is typically written as a disjunction $A_1 \vee A_2 \vee ... \vee A_n$ of \textit{actions} $A_i$, where an action is a logical predicate that depends on both the current and next state of a behavior. 
Correctness properties and system specifications in TLA+ are both written as temporal logic formulas. This allows one to express notions of property satisfaction in a concise manner. We say that a specification $S$ \emph{satisfies} a property $P$ iff the formula $S \Rightarrow P$ is valid (i.e. true under all assignments). 


\subsubsection{The TLA+ Proof System}

The TLA+ proof system \cite{cousineau2012tla}, abbreviated as TLAPS, is an accompanying tool for the TLA+ language that allows one to write and mechanically check hierarchically structured proofs \cite{lamport1995write} in TLA+. Proofs consist of a series of statements that support the proof goal, which is the top level statement that must be proved. Each statement, in turn, must be proved either compositely using a nested structural proof, or as a leaf proof via a backend solver. TLAPS is independent of any particular SMT solver or theorem prover, and includes support for various backends e.g. Z3 \cite{de2008z3}, Isabelle \cite{wenzel2008isabelle}, and Zenon \cite{bonichon2007zenon}.

Figure \ref{fig:sample-tlaps} shows an example of a lemma and its proof in TLAPS. The \textsc{assume-prove} idiom treats the lemma as an implication. That is, if \textit{Conditions} hold, then \textit{Implication} must follow. Leaf statements are proved using the {\BY} statement, and can reference theorems and lemmas by name, operator definitions, and previous statements by label. Each structural proof must end with a {\QED} statement, closing the goal of either a nested or overall proof.

\begin{figure}
    \footnotesize
    \raggedright
    \begin{tla}
      LEMMA NameOfLemma ==
      ASSUME Conditions
      PROVE Implication
      PROOF
        <1>1. Statement1.1  BY DEF Conditions
        <1>2. Statement1.2
          <2>1. Statement 2.1  BY UsefulTheorem
          <2>2. Statement 2.2  BY <1>1, <2>1
          <2>.  QED BY <2>2
        <1>.  QED BY <1>1, <1>2
    \end{tla}
\begin{tlatex}
\@x{\@s{24.59} {\LEMMA} NameOfLemma \.{\defeq}}%
\@x{\@s{24.59} {\ASSUME} Conditions}%
\@x{\@s{24.59} {\PROVE} Implication}%
\@x{\@s{24.59} {\PROOF}}%
 \@x{\@s{32.8}\@pfstepnum{1}{1.}\  Statement1 . 1\@s{4.1} {\BY} {\DEF}
 Conditions}%
\@x{\@s{32.8}\@pfstepnum{1}{2.}\  Statement1 . 2}%
 \@x{\@s{41.0}\@pfstepnum{2}{1.}\  Statement 2 . 1\@s{4.1} {\BY}
 UsefulTheorem}%
 \@x{\@s{41.0}\@pfstepnum{2}{2.}\  Statement 2 . 2\@s{4.1}
 {\BY}\@pfstepnum{1}{1} ,\,\@pfstepnum{2}{1}\ }%
\@x{\@s{41.0}\@pfstepnum{2}{} .\@s{4.1} {\QED} {\BY}\@pfstepnum{2}{2}\ }%
 \@x{\@s{32.8}\@pfstepnum{1}{} .\@s{4.1} {\QED} {\BY}\@pfstepnum{1}{1}
 ,\,\@pfstepnum{1}{2}\ }%
\end{tlatex}
    \caption{Example of a hierarchically structured TLAPS proof.}
    \label{fig:sample-tlaps}
\end{figure}


\subsection{The MongoRaftReconfig TLA+ Specification}
\label{sec:mrr-tla-spec}

\begin{figure}
    \begin{align*}
        Ty&peOK \triangleq \\
        &log  & &\in [Server \rightarrow Seq(\mathbb{N})] \\
        &committed & &\in  2^{\mathbb{N} \times \mathbb{N}}  \\
        &term & &\in [Server \rightarrow \mathbb{N}] \\
        &state & &\in [Server \rightarrow \{Primary, Secondary\}] \\
        &config & &\in[Server \rightarrow 2^{Server}] \\
        &configVersion & &\in[Server \rightarrow \mathbb{N}]\\
        &configTerm & &\in [Server \rightarrow \mathbb{N}]
    \end{align*}
    \caption{The state variables of the \textit{MongoRaftReconfig} protocol and their corresponding types stated as a type correctness predicate in TLA+. The notation $[A \rightarrow B]$ represents the set of all functions from set $A$ to set $B$ and $Seq(S)$ represents the set of all sequences containing elements from the set $S$.}
    \label{fig:mrr-state-vars}
\end{figure}

A formal TLA+ specification of \textit{MongoRaftReconfig} was originally included in \cite{schultz2021design}, but was not discussed in detail. This same specification serves as the basis for the TLAPS proofs presented in this paper, so we give a brief overview of the specification here. The complete specification can be found in the \textit{MongoRaftReconfig.tla} file of the supplementary material provided with this paper \cite{zenodo-tlaps-safety-proof}.

The state variables of the specification and their types are shown in Figure \ref{fig:mrr-state-vars}. The initial states, next state relation, and specification definition of \textit{MongoRaftReconfig} are summarized in Figure \ref{fig:mrr-tla-next}. The operator $Quorums(m)$ is defined as the set of all majority quorums \cite{vukolic2013origin} for a given set of servers $m$. Reconfigurations are modeled by the $Reconfig(s, m)$ action, which represents a reconfiguration that occurs on primary server $s$ to a new configuration with member set $m \in 2^{Server}$. Configuration propagation is modeled by the $SendConfig(s, t)$ action, which represents the propagation of a configuration from server $s$ to server $t$. Elections are modeled by the action $BecomeLeader(s, Q)$, which represents the election of server $s$ by a set of voters $Q$. The action $UpdateTerms(s, t)$ propagates the term of a server $s$ to server $t$, if the term of $s$ is newer than $t$. The actions $ClientRequest(s)$, $GetEntries(s,t)$, $RollbackEntries(s,t)$, and $CommitEntry(s,Q)$ are responsible for log related actions that are conceptually unrelated to reconfiguration, so we do not discuss their details here. Their full definitions can be found in the specifications provided in the supplementary material \cite{zenodo-tlaps-safety-proof}.

Note that our specifications are written at a deliberately high level of abstraction, ignoring some lower level details of the protocol. In practice, we have found the abstraction level of our specifications most useful for understanding and communicating the essential behaviors and safety characteristics of the protocol, while also serving to make our automated verification and proof efforts more feasible. In the future, however, we believe it would be valuable to explore techniques for formally relating our abstract specifications to real world protocol implementations, with an aim of verifying whether a system implementation faithfully reflects our high level specifications \cite{bornholt2021using, hawblitzel2015ironfleet, 2020extrememodeling}.



\begin{figure}
    \input{tla-mongo-raft-reconfig-next.tex}
    \caption{Summary of the \emph{MongoRaftReconfig} TLA+ specification. The full specification consists of \statlocprotspec{} lines of TLA+ code, excluding comments, and can be found in the \textit{MongoRaftReconfig.tla} file of the supplementary material.}
    \label{fig:mrr-tla-next}
\end{figure}



\section{Verification Problem Statement}
\label{sec:verification-problem-statement}

In this paper we establish that the \textit{MongoRaftReconfig} protocol satisfies \textit{LeaderCompleteness} and \textit{StateMachineSafety}, which are two key, high level safety properties of both the MongoDB replication system and standard Raft. Informally, the \textit{LeaderCompleteness} property states  that if a log entry is committed in term $T$, then it is present in the log of any leader in term $T' > T$. It is stated more precisely in Definition \ref{def:leader-completeness}, where $committed \in \mathbb{N} \times \mathbb{N}$ refers to the set of committed log entries as $(index, term)$ pairs, and $InLog(i,t,s)$ is a predicate determining whether a log entry $(i,t)$ is contained in the log of server $s$. \textit{StateMachineSafety} states that if two log entries are committed at the same log index, these entries must be the same, and is stated formally as Definition \ref{def:state-machine-safety}.

\newcommand{\mrrspec}{MRRSpec}
\newcommand{\mrrinit}{MRRInit}
\newcommand{\mrrnext}{MRRNext}



\begin{definition}[Leader Completeness]
    \label{def:leader-completeness}
    \begin{align*}
        \forall s \in &Server : \forall (cindex, cterm) \in committed : \\
        &( state[s] = Primary \wedge cterm < term[s]) \Rightarrow\\ &InLog(cindex, cterm, s)
    \end{align*}
\end{definition}

\begin{definition}[State Machine Safety]
    \label{def:state-machine-safety}
    \begin{align*}
        \A (ind_i, t_i),&(ind_j, t_j) \in committed : \\
        &(ind_i = ind_j) \Rightarrow (t_i = t_j)
    \end{align*}
\end{definition}
Both \emph{LeaderCompleteness} and \emph{StateMachineSafety} are safety properties. More specifically, they are both invariants, meaning that they must hold in all reachable states of \textit{MongoRaftReconfig}. Thus, our verification goals can be stated formally as Theorems \ref{thm:spec-implies-lc} and \ref{thm:spec-implies-sms}, where $MRRSpec$ refers to the specification of \textit{MongoRaftReconfig} as given in Figure \ref{fig:mrr-tla-next}. 
\begin{theorem}[MRRImpliesLeaderCompleteness]
    \label{thm:spec-implies-lc}
    \begin{align*}
        \mrrspec \Rightarrow \square LeaderCompleteness
    \end{align*}
\end{theorem}
\begin{theorem}[MRRImpliesStateMachineSafety]
    \label{thm:spec-implies-sms}
    \begin{align*}
        \mrrspec \Rightarrow \square StateMachineSafety
    \end{align*}
\end{theorem}
Theorems \ref{thm:spec-implies-lc} and \ref{thm:spec-implies-sms} are the safety results established and formally verified in this paper, and they can be found in the \textit{MongoRaftReconfigProofs.tla} file of our supplementary material \cite{zenodo-tlaps-safety-proof}. The proofs of Theorems \ref{thm:spec-implies-lc} and \ref{thm:spec-implies-sms} are discussed in Section \ref{sec:tlaps-proof}. Both of these theorems are proved using the help of an inductive invariant, which we discuss next, in
Section \ref{sec:inductive-invariant}.

\section{The Inductive Invariant}
\label{sec:inductive-invariant}


\subsection{Background}

A standard method to establish an invariant $Inv$ is to find an {\em inductive invariant} that implies $Inv$~\cite{manna2012temporal}.
Formally, a state predicate $Inv$ is an invariant of a system $Spec$ if the following holds:
\begin{align}
    \label{eq:spec-implies-inv-example}
    Spec \Rightarrow \square Inv
\end{align}
Suppose that $Spec$ is of the form $Spec=Init \wedge \square[Next]_{vars}$, as in the case of \textit{MongoRaftReconfig}. Then, in order to establish Formula \ref{eq:spec-implies-inv-example}, it is sufficient to find a state predicate $Ind$ such that the following conditions hold:
\begin{align}
    \label{eq:initiation}
    &Init \Rightarrow Ind\\
    \label{eq:consecution}
    &Ind \wedge Next \Rightarrow Ind'\\
    \label{eq:sufficiency}
    &Ind \Rightarrow Inv
\end{align}
Conditions \ref{eq:initiation} and \ref{eq:consecution} are referred to as \textit{initiation} and \textit{consecution}, respectively, and they are sufficient to show that $Ind$ is an inductive invariant. Conditions \ref{eq:initiation}, \ref{eq:consecution}, and \ref{eq:sufficiency} are together sufficient to establish Formula \ref{eq:spec-implies-inv-example}. 


In our case, $Spec$ instantiates to $MRRSpec$ and we have two instances of $Inv$, namely, \emph{LeaderCompleteness} and \textit{StateMachineSafety}. In principle, we need to discover two distinct inductive invariants, one for \emph{LeaderCompleteness} and another one for \textit{StateMachineSafety}. In our case, the same inductive invariant turns out to be sufficient for both properties.

\subsection{Invariant Overview}
\label{sec:inv-structure}

The inductive invariant that we developed for \textit{MongoRaftReconfig} is referred to as $MRRInd$ and consists of 20 high level conjuncts, shown in Figure \ref{fig:inductive-invariant}. Its full definition is given in \statlocindinv{} lines of TLA+ code, and is provided in the \textit{MongoRaftReconfigIndInv.tla} file of our supplementary material \cite{zenodo-tlaps-safety-proof}. 



\newcommand{\ca}[1]{\textcolor{black}{#1}}
\newcommand{\cb}[1]{\textcolor{black}{#1}}
\newcommand{\cc}[1]{\textcolor{black}{#1}}
\newcommand{\cd}[1]{\textcolor{black}{#1}}

\let\oldwedge\wedge
\renewcommand{\wedge}{\oldwedge \,}

\begin{figure}
    \small
    \begin{align*}
        &MRRInd \triangleq\\
        T
        &\begin{cases}
            &\wedge TypeOK
        \end{cases}\\
        E_1
        &\begin{cases}
            &\wedge \ca{ElectionSafety}\\
            &\wedge \ca{PrimaryConfigTermEqualToCurrentTerm}\\
            &\wedge \ca{ConfigVersionAndTermUnique}\\
            &\wedge \ca{PrimaryInTermContainsNewestConfigOfTerm}\\
            &\wedge \ca{ActiveConfigsOverlap}\\
            &\wedge \ca{ActiveConfigsSafeAtTerms}\\
        \end{cases}\\
        L_1
        &\begin{cases}
            &\wedge \cb{LogEntryInTermImpliesConfigInTerm}\\
            &\wedge \cb{PrimaryHasEntriesItCreated}\\
            &\wedge \cb{LogMatching}\\
        \end{cases}\\
        L_2
        &\begin{cases}
            &\wedge \cb{PrimaryTermAtLeastAsLargeAsLogTerms}\\
            &\wedge \cb{TermsOfEntriesGrowMonotonically}\\
            &\wedge \cb{UniformLogEntriesInTerm}\\ 
        \end{cases}\\
        C_1
        &\begin{cases}
            &\wedge \cc{CommittedEntryIndexesAreNonZero}\\
            &\wedge \cc{CommittedTermMatchesEntry}\\
        \end{cases}\\
        C_2
        &\begin{cases}
            &\wedge \cd{LeaderCompleteness}\\
            &\wedge \cd{LogsLaterThanCommittedMustHaveCommitted}\\
            &\wedge \cd{ActiveConfigsOverlapWithCommittedEntry}\\
            &\wedge \cd{NewerConfigsDisableCommitsInOlderTerm}    
        \end{cases}\\
        N
        &\begin{cases}
            &\wedge \cd{ConfigsNonEmpty}    
        \end{cases}
    \end{align*}
    \caption{Our inductive invariant for \textit{MongoRaftReconfig}.}
    \label{fig:inductive-invariant}
\end{figure}

The inductive invariant, shown in Figure \ref{fig:inductive-invariant}, is composed of several conceptually distinct subcomponents. The first conjunct, $TypeOK$, establishes basic type-correctness constraints on the state variables of \textit{MongoRaftReconfig}. This is necessary in most cases when stating inductive invariants in TLA+, since it is an untyped formalism \cite{lamport1999should}. The full definition of $TypeOK$ is shown in Figure \ref{fig:mrr-state-vars}. The initial set of 6 conjuncts, labeled as $E_1$ in Figure \ref{fig:inductive-invariant}, along with $TypeOK$, is itself an inductive invariant, and it establishes the $ElectionSafety$ property, a key auxiliary invariant of the protocol that is needed to establish $LeaderCompleteness$. The conjuncts in group $L_1$ are a set of invariants related to logs of servers in the system, and they collectively establish $LogMatching$, another important auxiliary invariant. The $L_2$ group establishes a few additional log related invariants, which rely on previous conjuncts. In general, these log related conjuncts are not fundamentally related to dynamic reconfiguration, but are necessary to state precisely for a protocol that manages logs in a Raft like fashion. Group $C_1$ establishes some required, trivial aspects of the set of committed log entries.  The conjunct group $C_2$ establishes the high level $LeaderCompleteness$ property, by relating how configurations interact with the set of committed log entries present in the system. Finally, the last conjunct, labeled as $N$, asserts that every configuration is non empty i.e. it contains some servers. This is an auxiliary invariant that is helpful for proving other facts. 

\subsection{Discovering an Inductive Invariant}
\label{sec:discover-ii}

Discovering such an inductive invariant for a protocol of this complexity is non-trivial. To our knowledge, this is the first inductive invariant proposed for a dynamic reconfiguration protocol that is built on a Raft based replication system. The discovery of $MRRInd$ took approximately 1-2 human months of work and it involved repeated efforts of iteration and refinement. To aid in this discovery process, we leveraged a technique proposed in \cite{lamport2018using} that utilizes the TLC explicit state model checker \cite{tlcmodelchecker} to probabilistically verify candidate inductive invariants. If a candidate $Inv$ is not inductive, the TLC model checker can, with some probability, report a \textit{counterexample to induction}. A \textit{counterexample to induction} is a state transition $s \rightarrow t$ satisfying $MRRNext$, where $s$ satisfies $Inv$ and $t$ violates $Inv$. These counterexamples are helpful to understand why a candidate invariant fails to be inductive, and how it may need to be modified or strengthened further. This probabilistic method can only be used on finite protocol instances, and it does not provide a proof that an invariant is inductive. Nevertheless, the technique proved to be highly effective, as it helped us to discover an inductive invariant that we eventually proved formally correct using TLAPS, as discussed more in Section \ref{sec:tlaps-proof}. Furthermore, we did not discover any errors in our inductive invariant during the TLAPS proof process. 

Note that, although having a tool for finding counterexamples to induction is helpful for finding errors in candidate inductive invariants, it still does not provide much guidance in developing an inductive invariant from scratch. That is, it does not necessarily provide a systematic methodology for converging to a correct inductive invariant. Rather, development of our inductive invariant still required a large amount of creativity and human reasoning, largely driven by strong prior intuitions about the correctness of the protocol. For example, rather than aiming to develop the entire inductive invariant at once, we were able to develop it in components, based partially on human intuition about certain auxiliary lemmas that we knew must hold true of the overall protocol. For example, the $ElectionSafety$ and the $LogMatching$ invariants (shown in Figure \ref{fig:inductive-invariant}) are two such lemmas that we worked on establishing first, before moving on to discover the additional conjuncts needed to establish the $LeaderCompleteness$ property.

\section{TLAPS Proofs}
\label{sec:tlaps-proof}

%
%


In this section we present an overview of our formally verified safety proof of \textit{MongoRaftReconfig}, which was completed using TLAPS, the TLA+ proof system \cite{cousineau2012tla}.  Section \ref{sec:proof-ii} gives an overview of the proof that $MRRInd$ is an inductive invariant of \textit{MongoRaftReconfig}, and Section \ref{sec:proof-safety} describes how this fact is used to prove \textit{LeaderCompleteness} and \textit{StateMachineSafety}, which are the key, high level safety properties that were defined in Section \ref{sec:verification-problem-statement}.

%
%
\subsection{TLAPS Proof of Inductive Invariant}
\label{sec:proof-ii}

To establish that $MRRInd$ is an inductive invariant, we must prove that $MRRInd$ satisfies both the initiation (\ref{eq:initiation}) and consecution (\ref{eq:consecution}) conditions for \textit{MongoRaftReconfig}, as described in Section \ref{sec:inductive-invariant}.  This is captured in Lemma \ref{lemma:ind-is-inductive-inv}.
\begin{lemma}[$MRRInd$ is an inductive invariant]
  \label{lemma:ind-is-inductive-inv}
  \begin{align*}
    &MRRInit \Rightarrow MRRInd \tag{a} \label{eq:ind:initiation}\\
    &MRRInd \wedge MRRNext \Rightarrow MRRInd' \tag{b} \label{eq:ind:consecution}
  \end{align*}
\end{lemma}
Cases (\ref{eq:ind:initiation}) and (\ref{eq:ind:consecution}) of Lemma \ref{lemma:ind-is-inductive-inv} represent, respectively, initiation and consecution. The initiation case of Lemma \ref{lemma:ind-is-inductive-inv} follows in a straightforward manner from the definitions of $MRRInit$ and $MRRInd$.  Proving the consecution case of Lemma \ref{lemma:ind-is-inductive-inv}, however, is the most difficult and time consuming aspect of the verification efforts presented in this paper. At a high level, this proof consists of showing that, assuming $MRRInd$ holds in a current state, every transition of the protocol upholds $MRRInd$ in the next state. To break this verification problem into smaller steps, we decompose the proof first by each conjunct of $MRRInd$, and then we decompose by each protocol transition.


Specifically, consider the definition of $MRRInd$, which is composed of 20 conjuncts (as shown in Figure \ref{fig:inductive-invariant}):
\begin{align*}
    MRRInd \triangleq I_1 \wedge I_2 \wedge \dots \wedge I_{20}
\end{align*}
Our first decomposition step breaks down case (\ref{eq:ind:consecution}) of Lemma \ref{lemma:ind-is-inductive-inv} into the following, independent proof goals, one for each conjunct of $MRRInd$:
\begin{align}
    \label{eq:conjunct-split}
    \begin{split}
      &MRRInd \wedge MRRNext \Rightarrow I_1'\\
      &MRRInd \wedge MRRNext \Rightarrow I_2'\\
      &\phantom{MRRInd\,\,} \vdots\\
      &MRRInd \wedge MRRNext \Rightarrow I_{20}'
    \end{split}
\end{align}
Furthermore, $MRRNext$ is the disjunction of eight protocol actions (as shown in Figure \ref{fig:mrr-tla-next}):
\begin{align}
  MRRNext \triangleq A_1 \vee A_2 \vee \dots \vee A_8
\end{align}
So, we further decompose each goal of Statement \ref{eq:conjunct-split} into one case for each protocol action. That is, we decompose each goal $MRRInd \wedge MRRNext \Rightarrow  I_j'$ into the following proof goals:
\begin{align}
    \begin{split}
      &MRRInd \wedge A_1 \Rightarrow I_j'\\
      &MRRInd \wedge A_2 \Rightarrow I_j'\\
      &\phantom{MRRInd\,\,} \vdots \\
      &MRRInd \wedge A_8 \Rightarrow I_j'\\
    \end{split}
\end{align}
Our proof follows this methodology for every conjunct of $MRRInd$ and every action of $MRRNext$. This produces a set of proof goals whose size is the product of the number of protocol actions (8) and the number of invariant conjuncts (20), totaling $8*20 = 160$ proof goals. This decomposition allowed us to focus on proving one, small goal at a time, while incrementally building a library of reusable lemmas. 
The TLAPS proof of Lemma \ref{lemma:ind-is-inductive-inv} can be found in the \textit{MongoRaftReconfigProofs.tla} file of our supplementary material \cite{zenodo-tlaps-safety-proof}, while our library of lemmas can be found in the \textit{Lib.tla}, \textit{BasicQuorumsLib.tla}, and \textit{LeaderCompletenessLib.tla} files.


%
%
\subsection{TLAPS Proof of Safety}
\label{sec:proof-safety}

Lemma \ref{lemma:ind-is-inductive-inv} establishes that \textit{MRRInd} is an inductive invariant of \textit{MongoRaftReconfig}. In this section we provide an overview of our proofs for establishing that \textit{MongoRaftReconfig} satisfies \textit{LeaderCompletness} and \textit{StateMachineSafety} (Theorems \ref{thm:spec-implies-lc} and \ref{thm:spec-implies-sms}), which utilize $MRRInd$. We do this by establishing lemmas \ref{lemma:ind-implies-lc} and \ref{lemma:ind-implies-sms}, which, together with Lemma \ref{lemma:ind-is-inductive-inv}, are sufficient to establish Theorems \ref{thm:spec-implies-lc} and \ref{thm:spec-implies-sms}.
\begin{lemma}
  \label{lemma:ind-implies-lc}
  \begin{align*}
    MRRInd \Rightarrow LeaderCompleteness
  \end{align*}
\end{lemma}
\begin{lemma}[IndImpliesStateMachineSafety]
  \label{lemma:ind-implies-sms}
  \begin{align*}
    MRRInd \Rightarrow StateMachineSafety
  \end{align*}
\end{lemma}
\textit{LeaderCompleteness} is a conjunct of $MRRInd$ so the implication of Lemma \ref{lemma:ind-implies-lc} follows trivially. The proof of Lemma \ref{lemma:ind-implies-sms}, which can be found in the \textit{StateMachineSafetyLemmas.tla} file of our supplementary material \cite{zenodo-tlaps-safety-proof}, is not trivial and we present it in the following section as a concrete example of a TLAPS proof.

\subsection{Example of a TLAPS Proof}
\label{sec:tlaps-example}

In this section we present the proof of Lemma \ref{lemma:ind-implies-sms} to serve as an example of TLAPS. The proof relies on one additional lemma, stated as Lemma \ref{lemma:commits-are-log-entries}. The proof of Lemma \ref{lemma:commits-are-log-entries} is contained in the \textit{StateMachineSafetyLemmas.tla} file of the supplementary material \cite{zenodo-tlaps-safety-proof}. 
\begin{lemma}[CommitsAreLogEntries]
  \label{lemma:commits-are-log-entries}
  \begin{align*}
    MRR&Ind \Rightarrow \\
    \forall c \in &committed : \exists s \in Server : \\
    &InLog(c.entry, s)
  \end{align*}
\end{lemma}

The TLAPS proof of Lemma \ref{lemma:ind-implies-sms} is shown in Figure \ref{fig:tlaps-except-ind-implies-sms}. The proof uses the \textsc{assume-prove} idiom to show that \textit{MRRInd} implies \textit{StateMachineSafety}. By the definition of the \textit{StateMachineSafety} property, we can establish the proof goal given in \tlatexnum{1}{1}. Steps \tlatexnum{1}{2} and \tlatexnum{1}{3} assume that $c1$ and $c2$ are arbitrary committed entries that share the same index but are not identical, and {\PROVE \FALSE} {\OBVIOUS} establishes that these assumptions will lead to a contradiction. \tlatexnum{1}{4} is a composite proof that shows that $c1$ and $c2$ cannot share the same term. Steps \tlatexnum{1}{5} through \tlatexnum{1}{8} use Lemma \ref{lemma:commits-are-log-entries} to show that there exist servers $s1$ and $s2$ that respectively contain the committed entries $c1$ and $c2$ in their logs. The two cases \tlatexnum{1}{9} and \tlatexnum{1}{10} show that if either $c1$ or $c2$ has a larger term than the other, then we derive a contradiction as expected. Finally, it suffices to only consider cases \tlatexnum{1}{9} and \tlatexnum{1}{10} because of step \tlatexnum{1}{4}, and hence the proof is complete.

\subsection{Proof Statistics}

%
%
%
We now present some summary statistics about our TLAPS proof and its development to give a better sense of its scope, size, and difficulty. The entire TLAPS proof, including the statement of the inductive invariant and the protocol specification, consists of \statloctlapsproofs{} lines of TLA+ code, excluding comments. \statlocindinv{} of these lines are used for defining the inductive invariant and \statlocprotspec{} of these lines are used for specifying the \textit{MongoRaftReconfig} protocol. There are a total of \statnumproofthms{} top level theorems and \statnumprooflemmas{} formally stated lemmas. In terms of proof effort, we spent approximately 4 human-months on development of the TLAPS proof, which does not include the time to develop the inductive invariant described in Section \ref{sec:inductive-invariant}. Development of the inductive invariant took approximately an additional 1-2 human-months of work. For the TLAPS proof system to check the correctness of the completed proof from scratch it takes approximately 38 minutes on a 2020 Macbook Air using 8 Apple M1 CPU Cores. This computation time consists mostly of queries to an underlying backend solver e.g. Isabelle or an SMT solver. 

\subsection{Experience with TLAPS}

The hierarchical structure enforced by TLAPS led to well organized and generally readable proofs in our experience. Despite our overall positive experience, there two main shortcomings of TLAPS that we highlight below.

First, TLAPS does not offer much guidance when a backend solver fails on a leaf proof. In general, TLAPS does not distinguish between obligations that fail because they are false, versus obligations that are too difficult for the backend solvers. Second, we found that the TLAPS library did not always cater to our needs as conveniently as we hoped. For example, the \textit{MongoRaftReconfig} specification includes state variables that are represented as TLA+ sequences, which are indexed using $\mathbb{N} \setminus \{0\}$. While the TLAPS standard library has theorems for induction on $\mathbb{N}$ (\textit{NaturalsInduction.tla}), we were not able to find direct support for induction over the domain of sequences. Support for induction over the domain of sequences was not seamless, yet we were able to prove the desired theorem by tailoring parts of the library to our needs.


\subsection{Discussion}

Formally verifying safety properties for a large, real world distributed protocol is, in our experience, a very labor intensive task. Even if one has built up strong intuitions about correctness of a protocol, verification may take several months. Nevertheless, we believe that formal verification is of great value since, even for protocols that have been formally specified or model checked, design errors are still possible. For example, a safety bug in EPaxos \cite{moraru2013there}, a well known variant of the original Paxos protocol, was discovered several years after its initial publication \cite{sutra2020correctness}, even though EPaxos was accompanied by a TLA+ specification and manual safety proof in its original publication. Similarly, a bug in one of Raft's original reconfiguration protocols was also discovered after initial publication \cite{ongaro2015-membership-bug}.

Furthermore, developing a formal inductive invariant and safety proof often provides deeper insights into why a protocol is correct, which fully automated techniques like model checking, on their own, are often unable to provide. Gaining deeper, formalized understanding of why a protocol is correct is valuable both from a theoretical perspective and also for system designers and engineers who may implement these protocols with extensions, modifications, or optimizations.

%
%


\begin{figure}
    \footnotesize
    \raggedright
\begin{tla}
  LEMMA IndImpliesStateMachineSafety ==
ASSUME MRRInd
PROVE StateMachineSafety
PROOF
    <1>0. TypeOK BY DEF MRRInd
    <1>1. SUFFICES
          \A c1, c2 \in committed : (c1.entry[1] = c2.entry[1]) => (c1 = c2)
          BY DEF StateMachineSafety
    <1>2. TAKE c1, c2 \in committed
    <1>3. SUFFICES ASSUME c1.entry[1] = c2.entry[1], c1 # c2
          PROVE FALSE OBVIOUS
    <1>4. c1.term # c2.term
        <2>1. SUFFICES ASSUME c1.term = c2.term
              PROVE FALSE OBVIOUS
        <2>2. c1.entry[2] = c2.entry[2] BY <2>1 DEF MRRInd, CommittedTermMatchesEntry
        <2>3. c1.entry[1] = c2.entry[1] BY <1>3
        <2>4. c1 = c2 BY <1>0, <2>1, <2>2, <2>3, Z3 DEF TypeOK
        <2>. QED BY <1>3, <2>4
    <1>5. PICK s1 \in Server : InLog(c1.entry, s1) BY CommitsAreLogEntries
    <1>6. PICK s2 \in Server : InLog(c2.entry, s2) BY CommitsAreLogEntries
    <1>7. log[s1][c1.entry[1]] = c1.term BY <1>5 DEF MRRInd, CommittedTermMatchesEntry, InLog, TypeOK
    <1>8. log[s2][c2.entry[1]] = c2.term BY <1>6 DEF MRRInd, CommittedTermMatchesEntry, InLog, TypeOK
    <1>9. CASE c1.term > c2.term
        <2>1. \E i \in DOMAIN log[s1] : log[s1][i] = c1.term BY <1>5 DEF MRRInd, CommittedTermMatchesEntry, InLog, TypeOK
        <2>2. \E i \in DOMAIN log[s1] : log[s1][i] > c2.term BY <1>9, <2>1 DEF TypeOK
        <2>3. Len(log[s1]) >= c2.entry[1] /\ log[s1][c2.entry[1]] = c2.term
            <3>1. c2.term <= c2.term BY DEF MRRInd, TypeOK
            <3>. QED BY <1>5, <2>2, <3>1 DEF MRRInd, LogsLaterThanCommittedMustHaveCommitted, TypeOK
        <2>4. log[s1][c1.entry[1]] = c2.term BY <1>3, <2>3 DEF MRRInd, CommittedEntryIndexesAreNonZero, TypeOK
        <2>. QED BY <1>4, <1>7, <2>4 DEF TypeOK
    <1>10. CASE c1.term < c2.term
        <2>1. \E i \in DOMAIN log[s2] : log[s2][i] = c2.term BY <1>6 DEF MRRInd, CommittedTermMatchesEntry, InLog, TypeOK
        <2>2. \E i \in DOMAIN log[s2] : log[s2][i] > c1.term BY <1>10, <2>1 DEF TypeOK
        <2>3. Len(log[s2]) >= c1.entry[1] /\ log[s2][c1.entry[1]] = c1.term
            <3>1. c1.term <= c1.term BY DEF MRRInd, TypeOK
            <3>. QED BY <1>6, <2>2, <3>1 DEF MRRInd, LogsLaterThanCommittedMustHaveCommitted, TypeOK
        <2>4. log[s2][c2.entry[1]] = c1.term BY <1>3, <2>3 DEF MRRInd, CommittedEntryIndexesAreNonZero, TypeOK
        <2>. QED BY <1>4, <1>8, <2>4 DEF TypeOK
    <1>. QED BY <1>4, <1>9, <1>10 DEF MRRInd, TypeOK
\end{tla}
\begin{tlatex}
\footnotesize
\@x{\@s{8.2} {\LEMMA} IndImpliesStateMachineSafety \.{\defeq}}%
\@x{ \@s{8.2}{\ASSUME} MRRInd}%
\@x{ \@s{8.2}{\PROVE} StateMachineSafety}%
  \@x{\@s{16.4}\@pfstepnum{1}{0.}\  TypeOK {\ } {\BY} {\DEF} MRRInd}%
\@x{\@s{16.4}\@pfstepnum{1}{1.}\  {\SUFFICES} \A\, c1 ,\, c2 \.{\in} committed \.{:} }%
  \@x{\@s{30.00} ( c1 . entry [ 1 ] \.{=} c2 . entry [ 1 ] ) \.{\implies} ( c1 \.{=} c2 )}%
\@x{\@s{30.00} {\BY} {\DEF} StateMachineSafety}%
\@x{\@s{16.4}\@pfstepnum{1}{2.}\  {\TAKE} c1 ,\, c2 \.{\in} committed}%
  \@x{\@s{16.4}\@pfstepnum{1}{3.}\  {\SUFFICES} {\ASSUME} c1 . entry [ 1 ]
 \.{=} c2 . entry [ 1 ] ,\, c1 \.{\neq} c2}%
\@x{\@s{32.00} {\PROVE} {\FALSE} {\ }{\OBVIOUS}}%
\@x{\@s{16.4}\@pfstepnum{1}{4.}\  c1 . term \.{\neq} c2 . term}%
  \@x{\@s{32.8}\@pfstepnum{2}{1.}\  {\SUFFICES} {\ASSUME} {\ } c1 . term \.{=} c2 .term}%
  \@x{\@s{45.00} {\PROVE} {\FALSE} {\ }{\OBVIOUS}}%
 \@x{\@s{32.8}\@pfstepnum{2}{2.}\  c1 . entry [ 2 ] \.{=} c2 . entry [ 2 ]}%
  \@x{\@s{45.00} {\BY}\@pfstepnum{2}{1}\  {\DEF} MRRInd ,\,}%
  \@x{\@s{45.00} CommittedTermMatchesEntry}%
 \@x{\@s{32.8}\@pfstepnum{2}{3.}\  c1 . entry [ 1 ] \.{=} c2 . entry [ 1 ] {\ } {\BY}\@pfstepnum{1}{3}\ }%
  \@x{\@s{32.8}\@pfstepnum{2}{4.}\  c1 \.{=} c2}%
  \@x{\@s{45.00} {\BY}\@pfstepnum{1}{0}
 ,\,\@pfstepnum{2}{1} ,\,\@pfstepnum{2}{2} ,\,\@pfstepnum{2}{3} ,\, Z3 {\DEF}
 TypeOK}%
\@x{\@s{32.8}\@pfstepnum{2}{.} {\ }{\QED} {\BY}\@pfstepnum{1}{3}
 ,\,\@pfstepnum{2}{4}\ }%
 \@x{\@s{16.4}\@pfstepnum{1}{5.}\  {\PICK} s1 \.{\in} Server \.{:} InLog ( c1
  . entry ,\, s1 )}%
  \@x{\@s{30.00} {\BY} CommitsAreLogEntries}%
 \@x{\@s{16.4}\@pfstepnum{1}{6.}\  {\PICK} s2 \.{\in} Server \.{:} InLog ( c2
  . entry ,\, s2 )}%
  \@x{\@s{30.00} {\BY} CommitsAreLogEntries}%
 \@x{\@s{16.4}\@pfstepnum{1}{7.}\  log [ s1 ] [ c1 . entry [ 1 ] ] \.{=} c1 .
  term}%
  \@x{\@s{32.00} {\BY}\@pfstepnum{1}{5}\  {\DEF} MRRInd ,\, CommittedTermMatchesEntry ,\,}%
  \@x{\@s{32.00}InLog ,\, TypeOK}%
 \@x{\@s{16.4}\@pfstepnum{1}{8.}\  log [ s2 ] [ c2 . entry [ 1 ] ] \.{=} c2 .
  term}%
  \@x{\@s{32.00} {\BY}\@pfstepnum{1}{6}\  {\DEF} MRRInd ,\, CommittedTermMatchesEntry ,\,}%
  \@x{\@s{32.00}InLog ,\, TypeOK}%
\@x{\@s{16.4}\@pfstepnum{1}{9.} {\ }{\CASE} c1 . term \.{>} c2 . term}%
 \@x{\@s{32.8}\@pfstepnum{2}{1.}\  \E\, i \.{\in} {\DOMAIN} log [ s1 ] \.{:}
  log [ s1 ] [ i ] \.{=} c1 . term}%
  \@x{\@s{45.00} {\BY}\@pfstepnum{1}{5}\  {\DEF} MRRInd ,\,}%
  \@x{\@s{45.00}CommittedTermMatchesEntry ,\, InLog ,\, TypeOK}%
 \@x{\@s{32.8}\@pfstepnum{2}{2.}\  \E\, i \.{\in} {\DOMAIN} log [ s1 ] \.{:}
  log [ s1 ] [ i ] \.{>} c2 . term}%
  \@x{\@s{45.00} {\BY}\@pfstepnum{1}{9}
 ,\,\@pfstepnum{2}{1}\  {\DEF} TypeOK}%
  \@x{\@s{32.8}\@pfstepnum{2}{3.}\  Len ( log [ s1 ] ) \.{\geq} c2 . entry [ 1 ]}%
  \@x{\@s{61.00} \.{\land} log [ s1 ] [ c2 . entry [ 1 ] ] \.{=} c2 . term}%
  \@x{\@s{49.19}\@pfstepnum{3}{1.}\  c2 . term \.{\leq} c2 . term {\ }{\BY} {\DEF}
 MRRInd ,\, TypeOK}%
\@x{\@s{49.19}\@pfstepnum{3}{.} {\ }{\QED} {\BY}\@pfstepnum{1}{5}
 ,\,\@pfstepnum{2}{2} ,\,\@pfstepnum{3}{1}\  {\DEF} MRRInd  ,\, TypeOK, \,}%
 \@x{\@s{58.00}LogsLaterThanCommittedMustHaveCommitted}%
 \@x{\@s{32.8}\@pfstepnum{2}{4.}\  log [ s1 ] [ c1 . entry [ 1 ] ] \.{=} c2 .
 term}%
 \@x{\@s{45.00} {\BY}\@pfstepnum{1}{3} ,\,\@pfstepnum{2}{3}\  {\DEF} MRRInd ,\, TypeOK ,\,}%
 \@x{\@s{45.00} CommittedEntryIndexesAreNonZero}%
\@x{\@s{32.8}\@pfstepnum{2}{.} {\ }{\QED} {\BY}\@pfstepnum{1}{4}
 ,\,\@pfstepnum{1}{7} ,\,\@pfstepnum{2}{4}\  {\DEF} TypeOK}%
\@x{\@s{16.4}\@pfstepnum{1}{10.} {\ }{\CASE} c1 . term \.{<} c2 . term}%
 \@x{\@s{32.8}\@pfstepnum{2}{1.}\  \E\, i \.{\in} {\DOMAIN} log [ s2 ] \.{:}
 log [ s2 ] [ i ] \.{=} c2 . term}%
 \@x{\@s{45.00} {\BY}\@pfstepnum{1}{6}\  {\DEF} MRRInd ,\, InLog ,\, TypeOK ,\,}%
 \@x{\@s{45.00}  CommittedTermMatchesEntry}%
 \@x{\@s{32.8}\@pfstepnum{2}{2.}\  \E\, i \.{\in} {\DOMAIN} log [ s2 ] \.{:}
 log [ s2 ] [ i ] \.{>} c1 . term}%
 \@x{\@s{45.00} {\BY}\@pfstepnum{1}{10}
 ,\,\@pfstepnum{2}{1}\  {\DEF} TypeOK}%
 \@x{\@s{32.8}\@pfstepnum{2}{3.}\  Len ( log [ s2 ] ) \.{\geq} c1 . entry [ 1 ]}%
 \@x{\@s{61.00}\.{\land} log [ s2 ] [ c1 . entry [ 1 ] ] \.{=} c1 . term}%
 \@x{\@s{49.19}\@pfstepnum{3}{1.}\  c1 . term \.{\leq} c1 . term {\ }{\BY} {\DEF}
 MRRInd ,\, TypeOK}%
 \@x{\@s{49.19}\@pfstepnum{3}{.} {\ }{\QED} {\BY}\@pfstepnum{1}{6}
 ,\,\@pfstepnum{2}{2} ,\,\@pfstepnum{3}{1}\  {\DEF} MRRInd ,\, TypeOK ,\,}%
 \@x{\@s{58.00} LogsLaterThanCommittedMustHaveCommitted}%
 \@x{\@s{32.8}\@pfstepnum{2}{4.}\  log [ s2 ] [ c2 . entry [ 1 ] ] \.{=} c1 .
 term}%
 \@x{\@s{45.00} {\BY}\@pfstepnum{1}{3} ,\,\@pfstepnum{2}{3}\  {\DEF} MRRInd ,\, TypeOK ,\,}%
 \@x{\@s{45.00} CommittedEntryIndexesAreNonZero}%
 \@x{\@s{32.8}\@pfstepnum{2}{.} {\ }{\QED} {\BY}\@pfstepnum{1}{4}
 ,\,\@pfstepnum{1}{8} ,\,\@pfstepnum{2}{4}\  {\DEF} TypeOK}%
 \@x{\@s{16.4}\@pfstepnum{1}{.} {\ }{\QED} {\BY}\@pfstepnum{1}{4}
 ,\,\@pfstepnum{1}{9} ,\,\@pfstepnum{1}{10}\  {\DEF} MRRInd ,\, TypeOK}%
\end{tlatex}
      \caption{The TLAPS proof of Lemma \ref{lemma:ind-implies-sms}.}
      \label{fig:tlaps-except-ind-implies-sms}
\end{figure}

\section{Related Work}
\label{sec:related-work}

Previously, there have been a variety of distributed protocols formalized using TLAPS, including Classic Paxos \cite{chand2016formal}, Byzantine Paxos \cite{lamport2011byzantizing}, and the Pastry distributed hash table protocol \cite{rigorous-pastry}. 

The Raft protocol, upon initial publication, included a TLA+ formal specification of its static protocol, without dynamic reconfiguration \cite{OngaroDissertation2014}. Later, a formal verification of the safety properties of the static Raft protocol was completed using the Verdi framework for distributed systems verification \cite{woos2016planning}. The formal verification of static Raft in Verdi consisted of approximately 50,000 lines of Coq \cite{bertot2013interactive}, took around 18 months to develop, and consisted of 90 total invariants. In comparison, our proof consists of \statloctlapsproofs{} lines of TLA+ code. Note, however, that it is difficult to directly compare our work with \cite{woos2016planning} because (1) our TLA+ specifications are written at a higher level of abstraction, and (2) part of the work in \cite{woos2016planning} was aimed at producing a verified, runnable Raft implementation, which was not our goal. The work of \cite{woos2016planning} did not include a verification of Raft's dynamic reconfiguration protocols. To our knowledge, our work is the first formally verified safety proof for a reconfiguration protocol that integrates with a Raft based system. 

In general, developing formally verified proofs and inductive invariants for real world distributed protocols remains a challenging and non-trivial problem. In recent years, tools like Ivy \cite{padon2016ivy} have attempted to ease the burden of inductive invariant discovery and verification by taking an interactive approach to invariant development, and constraining the specification language for describing these systems so it falls into a decidable fragment of first order logic \cite{piskac2010deciding}. These restrictions, however, can place additional burden on the user in cases where a protocol or its invariants do not naturally fall into this decidable fragment \cite{padon2017paxos}. 

Recent work has built on top of the Ivy system in an attempt to automatically infer inductive invariants for distributed protocols, with varying degrees of success. Tools like IC3PO \cite{goel2021symmetry, goel2021towards}, SWISS \cite{hance2021finding}, and DistAI \cite{yao2021distai} represent the state of the art in automated inductive invariant discovery for distributed protocols. With some human guidance, they have  recently been able to scale to larger protocols like Paxos, but have not yet been applied to protocols like Raft.

Apalache \cite{konnov2019tlaplus} is a symbolic model checker for TLA+ specifications that has been developed in recent years  and can check inductiveness of protocol invariants for bounded parameters. It does not, however, currently have any procedures for automatic discovery of inductive invariants. In future it would be interesting to compare the effectiveness of using Apalache versus the probabilistic, TLC-based method for finding counterexamples to induction when debugging a candidate inductive invariant.


\section{Conclusions and Future Work}
\label{sec:conclusions}

In this paper we presented, to our knowledge, the first formal verification of a reconfiguration protocol for a Raft based replication system. We used TLA+ and TLAPS, the TLA+ proof system, to formalize and mechanically verify our inductive invariant and safety proofs. 

In future, we are interested in exploring ways to further automate the inductive invariant discovery process to the extent possible. Formal verification of liveness properties of \textit{MongoRaftReconfig} is another possible avenue for future efforts. In addition, we are interested in examining how the compositional structure of the protocol could be exploited to improve the inductive invariant discovery or TLAPS proof process.


\balance

\bibliographystyle{ACM-Reference-Format}
\bibliography{references}

\end{document}